\documentclass[twocolumn,tighten,times]{aastex62}
\usepackage{color}
\usepackage{multirow}
\usepackage{chngcntr}
\usepackage[utf8x]{inputenc}
\usepackage{perpage}
\usepackage{mathrsfs}
\usepackage{lipsum}
\usepackage{amsmath}
\newcommand{\Oiii}{[O\,{\sc iii}]}

\newcommand{\Sii}{[S\,{\sc ii}]}
\newcommand{\Nii}{[N\,{\sc ii}]}

\newcommand{\Oabund}{12+$\log$(O/H)}

\graphicspath{{./}{figures/}}

\received{\today}
\revised{tomorrow}
\accepted{the day after tomorrow}

\submitjournal{ApJ}

\shorttitle{Metallicities of star-forming UDGs}
\shortauthors{Rong et al.}

\begin{document}

\title{Lessons on Star-forming Ultra-diffuse Galaxies from The Stacked Spectra of Sloan Digital Sky Survey}

\correspondingauthor{Yu Rong}
\email{rongyuastrophysics@gmail.com}

\author[0000-0002-2204-6558]{Yu Rong}
\altaffiliation{FONDECYT Postdoctoral Fellow}
\affiliation{Institute of Astrophysics, Pontificia Universidad Cat\'olica de Chile, Av.~Vicu\~na Mackenna 4860, 7820436 Macul, Santiago, Chile}
%\affiliation{Chinese Academy of Sciences South America Center for Astronomy, National Astronomical Observatories, Chinese 
%Academy of Sciences, Beijing 100012, China}
%\affiliation{Key Laboratory for Computational Astrophysics, National Astronomical Observatories, Chinese Academy of Sciences, 20A Datun Road, Chaoyang District, Beijing 100012, China}
%\nocollaboration

%\author[0000-0002-2368-6469]{Fuyan Bian}
%\affiliation{Research School of Astronomy and Astrophysics, Australian National University, Canberra, ACT 2611, Australia}
%\affiliation{European Southern Observatory, Alonso de C\'ordova 3107, Casilla 19001, Vitacura, Santiago 19, Chile}

\author{Kai Zhu}
\affiliation{School of Astronomy and Space Science, University of Chinese Academy of Sciences, Beijing 100049, China}
\affiliation{National Astronomical Observatories, Chinese Academy of Sciences, 20A Datun Road, Chaoyang District, Beijing 100101, China}

\author[0000-0002-2368-6469]{Evelyn J. Johnston}
\altaffiliation{FONDECYT Postdoctoral Fellow}
\affiliation{Institute of Astrophysics, Pontificia Universidad Cat\'olica de Chile, Av.~Vicu\~na Mackenna 4860, 7820436 Macul, Santiago, Chile}

\author{Hong-Xin Zhang}
\affiliation{CAS Key Laboratory for Research in Galaxies and Cosmology, Department of Astronomy, University of Science and Technology of China, China}
\affiliation{School of Astronomy and Space Sciences, University of Science and Technology of China, Hefei, 230026, China}

\author{Tianwen Cao}
\affiliation{Chinese Academy of Sciences South America Center for Astronomy, National Astronomical Observatories, Chinese 
Academy of Sciences, Beijing 100012, China}
\affiliation{Key Laboratory of Optical Astronomy, National Astronomical Observatories, Chinese Academy of Sciences, Beijing 100101, China}
\affiliation{School of Astronomy and Space Science, University of Chinese Academy of Sciences, Beijing 100049, China}
\affiliation{Institute of Astrophysics, Pontificia Universidad Cat\'olica de Chile, Av.~Vicu\~na Mackenna 4860, 7820436 Macul, Santiago, Chile}

\author[0000-0003-0350-7061]{Thomas H.~Puzia}
\affiliation{Institute of Astrophysics, Pontificia Universidad Cat\'olica de Chile, Av.~Vicu\~na Mackenna 4860, 7820436 Macul, Santiago, Chile}

%\author{Qi Guo}
%\affiliation{Chinese Academy of Sciences South America Center for Astronomy, National Astronomical Observatories, Chinese 
%Academy of Sciences, Beijing 100012, China}

%\author[0000-0002-4002-861X]{Xiao-Yu Dong}
%\affiliation{Department of Physics and Astronomy, California State University, Northridge, California 91330, USA}

%\author{Zheng Zheng}
%\affiliation{Chinese Academy of Sciences South America Center for Astronomy, National Astronomical Observatories, Chinese 
%Academy of Sciences, Beijing 100012, China}

\author[0000-0002-8835-0739]{Gaspar Galaz}
\affiliation{Institute of Astrophysics, Pontificia Universidad Cat\'olica de Chile, Av.~Vicu\~na Mackenna 4860, 7820436 Macul, Santiago, Chile}

%\author{Mora Marcelo}
%\affiliation{Institute of Astrophysics, Pontificia Universidad Cat\'olica de Chile, Av.~Vicu\~na Mackenna 4860, 7820436 Macul, Santiago, Chile}

%\author{Giuseppe D'Ago}
%\affiliation{Institute of Astrophysics, Pontificia Universidad Cat\'olica de Chile, Av.~Vicu\~na Mackenna 4860, 7820436 Macul, Santiago, Chile}

%\author{more collaborators}

%%%%%%%%%%%%%%%%%%%%%%%%%%%%%%%%%%%%%%%%
\begin{abstract}
	
	We investigate the on-average properties for 28 star-forming ultra-diffuse galaxies (UDGs) located in low-density environments, by stacking their spectra from the Sloan Digital Sky Survey. These relatively-isolated UDGs, with stellar masses of $\log_{10}(M_*/M_{\odot})\sim 8.57\pm0.29$, have the on-average total-stellar-metallicity [M/H]$\sim -0.82\pm0.14$, iron-metallicity [Fe/H]$\sim -1.00\pm0.16$, stellar age $t_*\sim5.2\pm0.5$~Gyr, $\alpha$-enhancement [$\alpha$/Fe]$\sim 0.24\pm0.10$, and oxygen abundance \Oabund$\sim 8.16\pm0.06$, as well as central stellar velocity dispersion $54\pm12$~km/s. On the star-formation rate versus stellar mass diagram, these UDGs are located lower than the extrapolated star-forming main sequence from the massive spirals, but roughly follow the main sequence of low-surface-brightness dwarf galaxies. We find that these star-forming UDGs are not particularly metal-poor or metal-rich for their stellar masses, as compared with the metallicity-mass relations of the nearby typical dwarfs. With the UDG data of this work and previous studies, we also find a coarse correlation between [Fe/H] and magnesium-element enhancement [Mg/Fe] for UDGs: [Mg/Fe]$\simeq-0.43(\pm0.26)$[Fe/H]$-0.14(\pm0.40)$.
	
\end{abstract}

\keywords{galaxies: dwarf --- galaxies: evolution --- methods: statistical --- galaxies: stellar content}

%%%%%%%%%%%%%%%%%%%%%%%%%%%%%%%%%%%%%%%%%

\section{Introduction} \label{sec:intro}

As a possible challenge to current galaxy formation models, many properties of the population of ultra-diffuse galaxies \citep[UDGs;][]{vanDokkum15,Mihos15}, including but not limited to their halo masses and dark matter fractions \citep[e.g.,][]{vanDokkum16,vanDokkum18}, spins \citep{Rong17a,Leisman17}, alignments and morphologies \citep[e.g.,][]{Yagi16,Rong19a,Rong19b,Rong20}, gas content and star formation \citep[e.g.,][]{Trujillo17,Leisman17}, and particularly, metallicities \citep[e.g.,][]{Gu18,Ruiz-Lara18,Ferre-Mateu18,Pandya18,Martin-Navarro19,Fensch19}, are still not clear. To the present, the studies for UDG metallicities are almost focused on the quiescent members in galaxy clusters/groups; the metallicity properties of star-forming UDGs are barely investigated.

Metallicity is one of the fundamental observational quantities that could provide information about the evolution of UDGs. The metal content of a galaxy is determined by a complex interplay between cosmological gas inflow, metal production by stars, and gas outflow via feedback. Inflows usually dilute the metallicity of a galaxy \citep[e.g.,][]{Rupke10} while provide fuel for star formation, which then convert hydrogen and helium to heavier elements. The outflows driven by stellar or AGN feedback inject energy into the interstellar medium and flow the gas and metals out of the galaxy \citep[e.g.,][]{Rong17b,Christensen18}. The ejected metals can escape from the gravitational potential well of the galaxy or be re-accreted into the galaxy and enrich it again. Measuring the gas-phase and stellar metallicities thus augments the understanding of the importance of outflows/inflows during UDG formation. The studies of the $\alpha$-element enhancement of UDG stellar population can, however, provide clues about the time-scale of star formation in UDGs. The $\alpha$-enhancement is measured through the [$\alpha$/Fe] ratio, where $\alpha$-elements and irons (Fe) are ejected into the interstellar medium primarily by Type~II and Ia supernovae (SN~II and SN~Ia), respectively. Since SN~Ia start to occur $\sim1$~Gyr after the onset of star formation while SN~II appear much sooner, the ratio of $\alpha$ elements, such as magnesium to iron ([Mg/Fe]), can be used to estimate relative star-formation timescales. A shorter episode of star formation in a UDG will result in an $\alpha$-enhanced stellar population due to the enrichment of magnesium from the SN~II, and the $\alpha$-enhancement will begin to drop after SN~Ia appear due to the dilution of magnesium with iron in the interstellar medium \citep{Thomas05}.

%In addition, the abundance of carbon monoxide (CO), a well-known tracer of the molecular gas in galaxies, is also related to galaxy metallicity; the CO-to-H$_2$ conversion factors ($\alpha_{\rm{CO}}$) of low-surface-brightness galaxies are known to be usually much higher than that of the Milky Way, due primarily to the lack of metals and dust grains \citep{Glover11,Feldmann12}. Therefore, the studies of metallicities in the low-surface-brightness population, UDGs, are virtual for the estimate of integration time for the follow-up detection of CO emission in UDGs with submillimeter telescopes.

We will select a sample of star-forming UDGs with spectra from the Sloan Digital Sky Survey (SDSS), located in the low-density environments, and stack their spectra to obtain a relatively-high signal-to-noise ratio (S/N) spectrum, and then study the on-average metallicity with the stacked spectrum. In section~\ref{sec:2}, we will describe the selection of UDG sample. We will describe the method of stacking the spectra of UDGs and investigate the on-average UDG properties in section~\ref{sec:3}, as well as discuss our results in section~\ref{sec:4}. In this paper, we assume the Hubble constant $H_0=69.6$~km/s \citep{Bennett14}, and use ``log'' to represent ``$\log_{10}$''.

%%%%%%%%%%%%%%%%%%%%%%%%%%%%%%%%%%%%%%%%%%%%%
\section{Selecting UDGs in SDSS}\label{sec:2}

\begin{figure}[!]
\centering
\includegraphics[angle=0,width=\columnwidth]{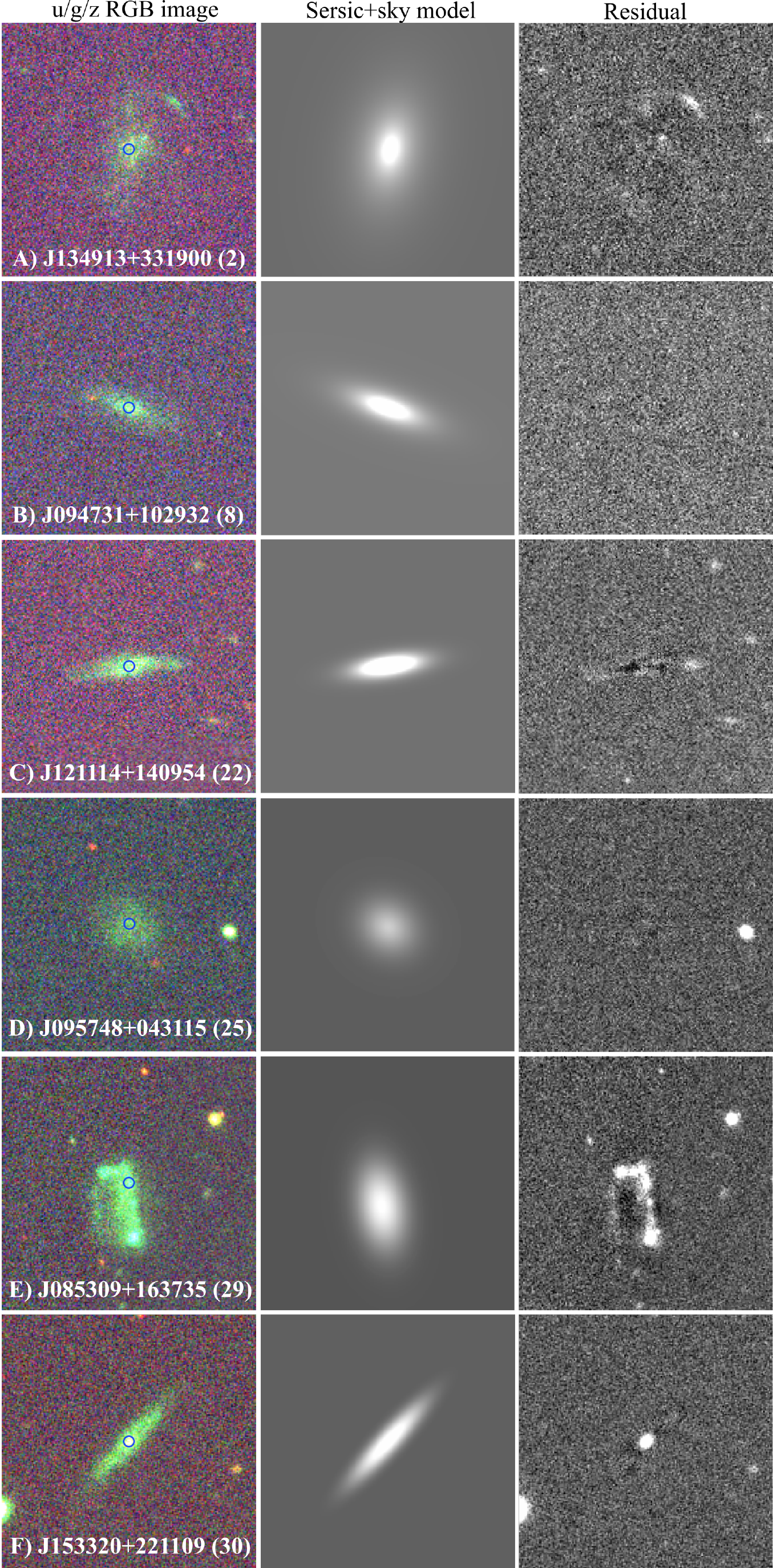}
\caption{The left, middle, and right panels show the original RGB images, fitting models, and residuals ($g$-band), respectively. Panels~A,~B,~C, and~D show the examples of the selected UDGs in this work; panels~E and~F exhibit the two abandoned UDG candidates since the SDSS 3~arcsec spectroscopic fiber (blue circles in the left panels show the 3~arcsec aperture) targets at the star-forming region and central nucleus with the distinct color from the entire stellar body, respectively. The SDSS names of these galaxies are also shown in the corresponding panels; the numbers in the brackets correspond to the UDG numbers in Table~\ref{UDG}.}
\label{image_fit}
\end{figure}

\begin{figure*}[!]
\centering
\includegraphics[angle=0,width=\textwidth]{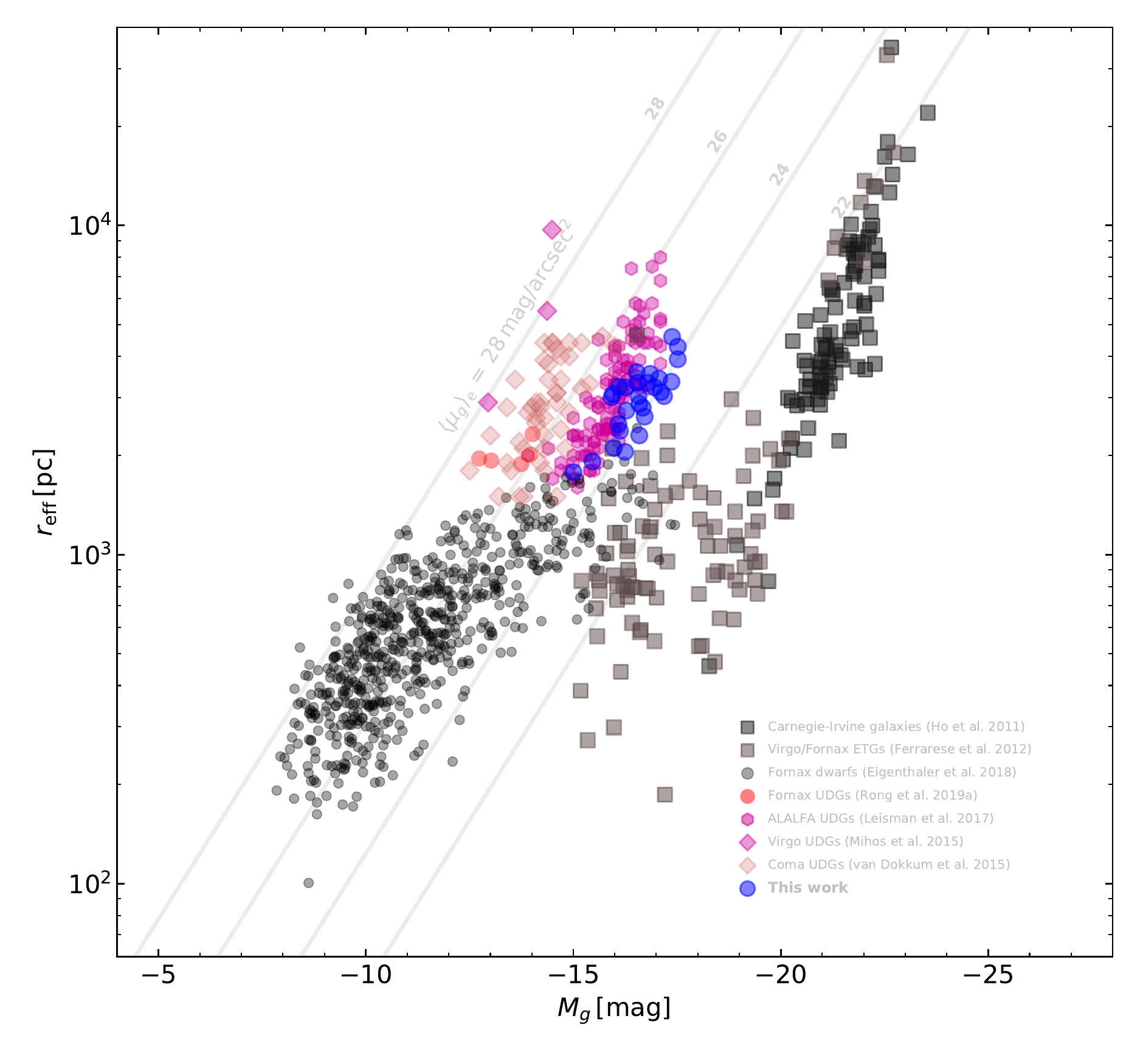}
\caption{The scale relation for the UDGs in this work (blue filled circles), and UDGs in the Virgo \citep[purple diamonds;][]{Mihos15}, Coma \citep[light-orange diamonds;][]{vanDokkum15}, and Fornax \citep[red filled circles;][]{Rong19a} clusters, as well as fields \citep[purple hexagons;][]{Leisman17}, plotted on that of typical dwarfs (black filled circles) and massive galaxies (squares) in the nearby clusters \citep{Ho11,Ferrarese12}.}
\label{scalerelation}
\end{figure*}

We first select a sample of low-surface-brightness galaxies with the mean surface brightness (within effective radius $r_{\rm{eff}}$) $\langle\mu_{\rm{eff,abs}}(r)\rangle> 22.5\ \rm{mag/arcsec^2}$ from the galaxy catalog of \cite{Simard11}, which contains 670,131 galaxies with SDSS optical spectra; each galaxy was roughly fitted with a pure S\'ersic model by \cite{Simard11}; we only select the large galaxies with $r_{\rm{eff}}> 1.5$~kpc as the candidates. The optical images of these candidates are then inspected by eyes to further abandon the objects being the substructures of large galaxies or having close companions such as bright stars or galaxies. 103 candidates are preliminarily selected.

For each selected candidate, we utilize a S\'ersic+sky model to fit its $g$ and $r$-band fully-processed SDSS images with \textsc{Galfit} \citep{Peng10}, by using the iterative fitting methodology outlined in \cite{Eigenthaler18} (to remove the contaminations of member globular clusters, background interlopers, and star-forming regions, etc). In Fig.~\ref{image_fit}, we show the fitting results of several examples. %Following the UDG definition described in \cite{Rong17a}, only t
The stellar masses are estimated by using the $r$-band luminosities and stellar mass-to-light ratios derived from the Galactic extinction corrected colors, $\log(M_*/L)=-0.306-0.15+1.097\times(g-r)$ \citep{Guo20}. The 33 galaxies with the $g$-band central surface brightness $\mu_{0,g}>23.5\ \rm{mag/arcsec^2}$, $r_{\rm{eff}}> 1.5$~kpc, and $\log M_*<9.0$ are selected as UDGs.

Among the 33 UDGs, there are 5 UDGs for which the SDSS $3''$ fiber aperture targeted at their star-forming regions (e.g., panel~E of Fig.~\ref{image_fit}) or central nuclei/small-bulges (e.g., panel~F of Fig.~\ref{image_fit}); these regions exhibit the colors significantly different from the colors of their entire stellar bodies in their RGB images. Therefore, the 5 UDGs are further abandoned, since their SDSS spectra cannot reveal the on-average properties of these UDGs. Finally, only 28 UDGs with SDSS spectra are selected, as listed in Table~\ref{UDG}; each selected UDG resides in the low-density environment, i.e., outside of the virial radius ($R_{\rm{vir}}$) of the nearest galaxy group/cluster \citep{Saulder16}. As explored in Fig.~\ref{scalerelation}, our UDG sample represents the relatively-bright UDG population; it is because that our UDGs are star-forming (cf.~Fig.~\ref{spectra}), and thus have relatively-lower $M_*/L$ than those UDGs in clusters; therefore, the same $M_*$ range corresponds to relatively-brighter star-forming UDGs.

\begin{table*}[!] 
\begin{center}
\begin{tabular}{@{}lccccccccccccc@{}}
\hline
\hline
Num & RA & DEC & $z$ & distance & $\mu_{0,g}$ & $r_{\rm{eff}}$ & $m_{r,\rm{f}}$ & $M_r$ & $g-r$ & $\log M_*$ & R/R$_{\rm{vir}}$ & SFR$_{\rm{fiber}}$ & SFR$_{\rm{tot}}$ \\
 & (deg) & (deg) &  & (Mpc) & (mag/$''^2$) & (kpc) & (mag) & (mag) & (mag) & ($\log M_{\odot}$) &  & ($M_{\odot}$/yr) & ($M_{\odot}$/yr) \\
\hline
1 & 164.087  & 56.760  & 0.00615  & 29.6  & 23.75  & 2.9  &  19.49  & -17.14  & 0.56   & 8.9   & 1.83 &  $1.0\times10^{-4}$ & $6.1\times10^{-3}$ \\
2 & 207.304  & 33.317  & 0.00723  & 35.0  & 23.75  & 1.8  &  20.95  & -15.32  & 0.31   & 7.9   & 5.37 &  $1.9\times10^{-4}$ & $6.4\times10^{-3}$ \\
3 & 232.685  & 47.319  & 0.00855  & 38.0  & 24.02  & 3.6  &  20.27  & -17.07  & 0.54   & 8.8   & 2.93 &  $3.0\times10^{-3}$ & $1.4\times10^{-1}$ \\
4 & 185.314  & 58.085  & 0.00944  & 41.3  & 23.51  & 2.3  &  19.50  & -17.02  & 0.43   & 8.7   & 8.60 &  $8.9\times10^{-4}$ & $2.3\times10^{-2}$ \\
5 & 234.388  & 58.580  & 0.00972  & 42.7  & 23.85  & 2.4  &  19.72  & -16.58  & 0.45   & 8.5   & 1.28 &  $1.8\times10^{-3}$ & $3.2\times10^{-2}$ \\
6 & 111.809  & 42.204  & 0.01003  & 44.8  & 24.53  & 3.1  &  21.01  & -16.46  & 0.51   & 8.5   & 3.45 &  $8.4\times10^{-4}$ & $3.1\times10^{-2}$ \\
7 & 234.284  & 20.146  & 0.01027  & 45.9  & 25.02  & 3.0  &  20.82  & -16.19  & 0.26   & 8.2   & 4.96 &  $8.2\times10^{-3}$ & $1.6\times10^{-1}$ \\
8 & 146.881  & 10.492  & 0.01044  & 49.6  & 23.97  & 1.9  &  20.41  & -15.75  & 0.29   & 8.0   & 1.74 &  $1.3\times10^{-3}$ & $2.0\times10^{-2}$ \\
9 & 177.654 & 24.926  & 0.01216  & 56.7  & 23.86  & 2.5  &  20.39  & -16.35  & 0.27   & 8.2   & 1.80 &   $4.7\times10^{-4}$ & $8.1\times10^{-3}$ \\
10 & 146.339 & 14.580  & 0.01267  & 59.3  & 24.36  & 3.2  &  19.96  & -16.59  & 0.33   & 8.4   & 5.75 &  $2.7\times10^{-3}$ & $3.5\times10^{-2}$ \\
11 & 139.232 & 14.714  & 0.01314  & 58.5  & 24.15  & 2.7  &  20.76  & -16.64  & 0.36   & 8.4   & 8.37 &  $2.8\times10^{-4}$ & $7.7\times10^{-3}$ \\
12 & 48.454  & -8.147  & 0.01372  & 56.8  & 23.68  & 2.8  &  19.66  & -16.90  & 0.23   & 8.4   & 2.76 &  $9.6\times10^{-3}$ & $9.0\times10^{-2}$ \\
13 & 187.568 &  3.073  & 0.01366  & 63.9  & 23.66  & 2.1  &  19.93  & -16.26  & 0.28   & 8.2   & 10.5 &  $2.6\times10^{-3}$ & $2.3\times10^{-2}$ \\
14 & 191.489 & 35.171  & 0.01453  & 68.2  & 24.00  & 3.3  &  19.95  & -16.87  & 0.31   & 8.5   & 6.24 &  $2.3\times10^{-3}$ & $3.0\times10^{-2}$ \\
15 & 157.110 & 31.262  & 0.01497  & 69.0  & 23.56  & 2.0  &  20.26  & -16.50  & 0.25   & 8.3   & 6.06 &  $3.3\times10^{-3}$ & $3.5\times10^{-2}$ \\
16 & 153.282 & 36.096  & 0.01491  & 69.0  & 24.72  & 3.2  &  20.41  & -16.48  & 0.38   & 8.4   & 7.84 &  $1.3\times10^{-3}$ & $1.7\times10^{-2}$ \\
17 & 240.561 & 17.506  & 0.01589  & 70.4  & 24.00  & 3.4  &  20.29  & -17.42  & 0.35   & 8.8   & 2.98 &  $1.4\times10^{-3}$ & $3.2\times10^{-2}$ \\
18 & 202.486 & -0.614  & 0.01652  & 75.7  & 23.65  & 3.0  &  19.91  & -17.54  & 0.35   & 8.8   & 6.39 &  $2.5\times10^{-3}$ & $3.6\times10^{-2}$ \\
19 & 121.914 & 56.925  & 0.01832  & 80.7  & 23.68  & 3.0  &  20.67  & -17.02  & 0.45   & 8.7   & 6.41 &  $1.8\times10^{-3}$ & $3.1\times10^{-2}$ \\
20 & 152.880 & 65.090  & 0.02008  & 88.1  & 23.83  & 3.9  &  19.32  & -17.86  & 0.34   & 8.9   & 13.7 &  $1.3\times10^{-2}$ & $1.2\times10^{-1}$ \\
21 & 197.298 & 28.777  & 0.02123  & 94.6  & 23.84  & 2.6  &  20.70  & -17.06  & 0.34   & 8.6   & 2.28 &  $3.5\times10^{-3}$ & $4.5\times10^{-2}$ \\
22 & 182.807 & 14.165  & 0.02119  & 96.2  & 24.27  & 3.5  &  20.13  & -17.31  & 0.46   & 8.8   & 6.73 &  $7.6\times10^{-3}$ & $6.3\times10^{-2}$ \\
23 & 171.122 & 34.581  & 0.02128  & 96.6  & 23.88  & 3.2  &  19.83  & -17.34  & 0.39   & 8.8   & 1.80 &  $4.8\times10^{-3}$ & $4.1\times10^{-2}$ \\
24 & 176.358 & 32.252  & 0.02150  & 97.2  & 23.97  & 3.3  &  20.13  & -17.18  & 0.39   & 8.7   & 5.47 &  $3.9\times10^{-3}$ & $3.5\times10^{-2}$ \\
25 & 149.448 &  4.521  & 0.02137  & 99.4  & 23.90  & 3.1  &  20.84  & -17.55  & 0.43   & 8.9   & 9.87 &  $1.3\times10^{-3}$ & $2.8\times10^{-2}$ \\
26 & 177.310 & 36.763  & 0.02196  & 98.4  & 24.29  & 4.6  &  19.87  & -17.66  & 0.28   & 8.8   & 6.77 &  $5.7\times10^{-3}$ & $6.5\times10^{-2}$ \\
27 & 256.171 & 62.035  & 0.02324  & 99.8  & 23.75  & 4.3  &  19.84  & -17.85  & 0.33   & 8.9   & 15.8 &  $1.4\times10^{-2}$ & $1.4\times10^{-1}$ \\
28 & 166.879 & 16.755  & 0.02665  & 120.0 & 23.57  & 3.4  &  19.93  & -17.72  & 0.35   & 8.9   & 10.2 &  $5.2\times10^{-3}$ & $4.5\times10^{-2}$ \\
\hline
\hline
\end{tabular}
\end{center}
\caption{Properties of the selected UDGs.  
Col. (1): galaxy number; Col. (2): RA; Col. (3): DEC; 
Col. (4): spectroscopic redshift; Col. (5): distance to us (corrected for CMB), obtained from the NASA/IPAC Extragalactic Database; 
Col. (6): observed $g$-band central surface brightness; Col. (7): $g$-band effective radius; Col. (8): $r$-band SDSS fiberMag without correction for extinction; Col. (9): $r$-band absolute magnitude corrected for Galactic extinction; Col. (10): $g-r$ color corrected for Galactic extinction; Col. (11): estimated stellar mass; Col. (12): three-dimensional distance to the nearest galaxy cluster/group, normalized by the virial radius of group/cluster; Col. (13): SFR covered by the SDSS $3''$ fiber aperture; Col. (14): estimated total star-formation rate.
}
\label{UDG}
\end{table*}

%%%%%%%%%%%%%%%%%%%%%%%%%%%%%%%%%%%%%%%%%%%%%
\section{Data analysis and UDG properties}\label{sec:3}

\begin{figure*}[!]
\centering
\includegraphics[angle=0,width=\textwidth]{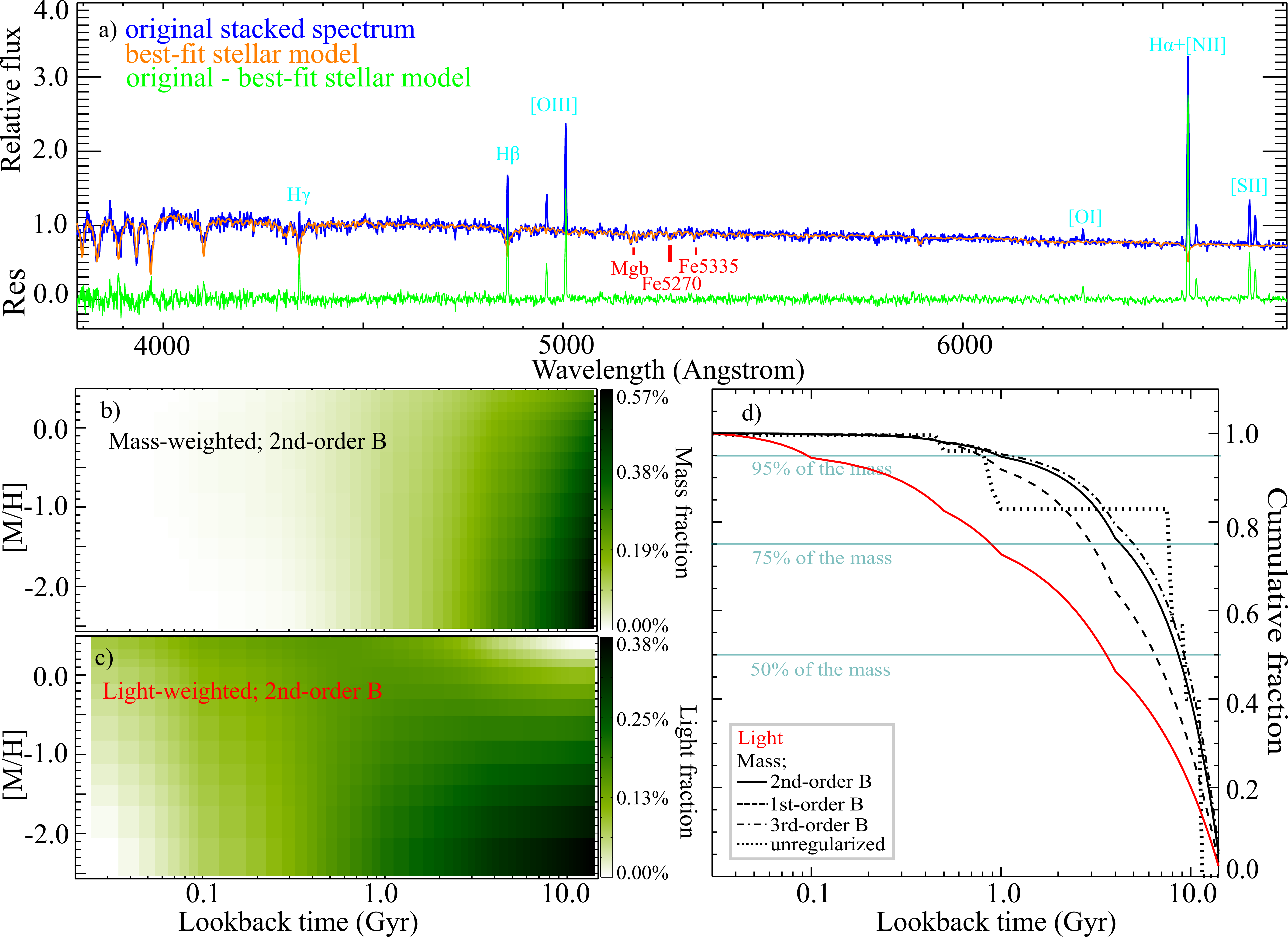}
\caption{Panel~a: the stacked spectrum (blue), best-fit stellar model (orange), and residual (green) of the selected 28~UDGs. The Mg-feature, Mgb, and Fe-features, Fe5270 \& Fe5335, are also shown. Panels~b \& c show the regularized (with the 2nd-order regularization matrix) mass-weighted and light-weighted stellar age-metallicity distributions revealing SFH, respectively, with the color-bars giving the mass and light fractions corresponding to each value for age and metallicity. Panel~d: the mass assembly (black) and light assembly (red) of our UDGs; the dark-green lines indicate when our UDGs had already formed 50\%, 75\%, and 95\% of their stellar masses/light, respectively. For comparison, we show the SFH without regularization (dotted) as well as SFHs with the regularized solutions of applying the first- (dashed), second- (solid), and third-order (dot-dashed) regularization matrix $\bm B$, i.e., $\bm B$=diag(1,-1), $\bm B$=diag(1,-2,1), and $\bm B$=diag(1,-3,3,-1), respectively \citep[cf.][]{Boecker20}.}
\label{spectra}
\end{figure*}

Since the S/N (defined as the median S/N in 5490\--5510~\AA) of an individual UDG spectrum is low (S/N$\sim 3\--14$), we stack the spectra of the selected UDGs and study the on-average stellar and gas-phase metallicities. For each galaxy spectrum from SDSS, 
we first correct it for the Galactic extinction by using the extinction curve of \cite{Fitzpatrick99} with $R_V=3.1$ and $E(B-V)$ value from the NASA/IPAC Extragalactic Database; the spectrum is then shifted to the rest frame and interpolated onto a wavelength grid spanning $3790\-–6800\AA$ with spacing $\Delta \ln \lambda\ (\AA)=1$. Each spectrum is normalized with the median flux density in $4400\--4450\AA$. We then stack the spectra using the median flux density at each wavelength (S/N$\sim 30$ for the stacked spectrum); the stacked spectrum is shown in Fig.~\ref{spectra}. The significant H$\alpha$ emission line indicates that our relatively-isolated UDGs are star-forming.

Since the old stellar population would be shaded by the light of the recently-formed stars in our star-forming UDGs, in order to study the star-formation history (SFH) and mass-weighted properties for our UDGs, analogous to the work of \cite{Fahrion19} and \cite{Rong18a}, we use \textsc{pPXF} \citep[V7.3.0]{Cappellari17} to fit the stacked spectrum, with the MILES single stellar population (SSP) template spectra \citep{Vazdekis15}, plus emission-line models (assuming the Balmer decrement for Case B recombination). The MILES models implement the BaSTI isochrones \citep{Pietrinferni06} and a Milky Way-like, double power law (bimodal), initial mass function (IMF) with a high mass slope of 1.30, and include 53 ages from 30~Myr to 14~Gyr, and 12 stellar metallicities from [M/H]=-2.27 to +0.40. We follow the linear regularization process of \cite[adopting the second-order regularization matrix, i.e., the \textsc{pPXF} option `REG\_ORD'=2]{McDermid15} to smooth the variation in the weights of templates of similar ages and metallicities. Since the original MILES library only offers the scaled solar models ([$\alpha$/Fe]=0) and alpha enhanced models ([$\alpha$/Fe]=0.4~dex), using a regularized \textsc{pPXF} solution seems unphysical; to develop a better sampled grid of SSP models for the fits, we linearly interpolate between the available SSPs to create a grid from [$\alpha$/Fe]=0 to [$\alpha$/Fe]=0.4~dex with a spacing of 0.1~dex, following the same method described in \cite{Fahrion19}. These models are created under the assumption that the [$\alpha$/Fe] abundances behave linearly in this regime and only give the average [$\alpha$/Fe]; however, note that in reality the abundances of different $\alpha$-elements might be decoupled. These $\alpha$-variable MILES models allow to study the distribution of $\alpha$-abundances from high S/N spectra. We set up \textsc{pPXF} to use the multiplicative polynomials of the 10th order, and derive the optimal (best-fit) stellar template. The best-fit stellar spectrum continuum is shown in Fig.~\ref{spectra}.

\begin{figure*}[!]
\centering
\includegraphics[width=\textwidth]{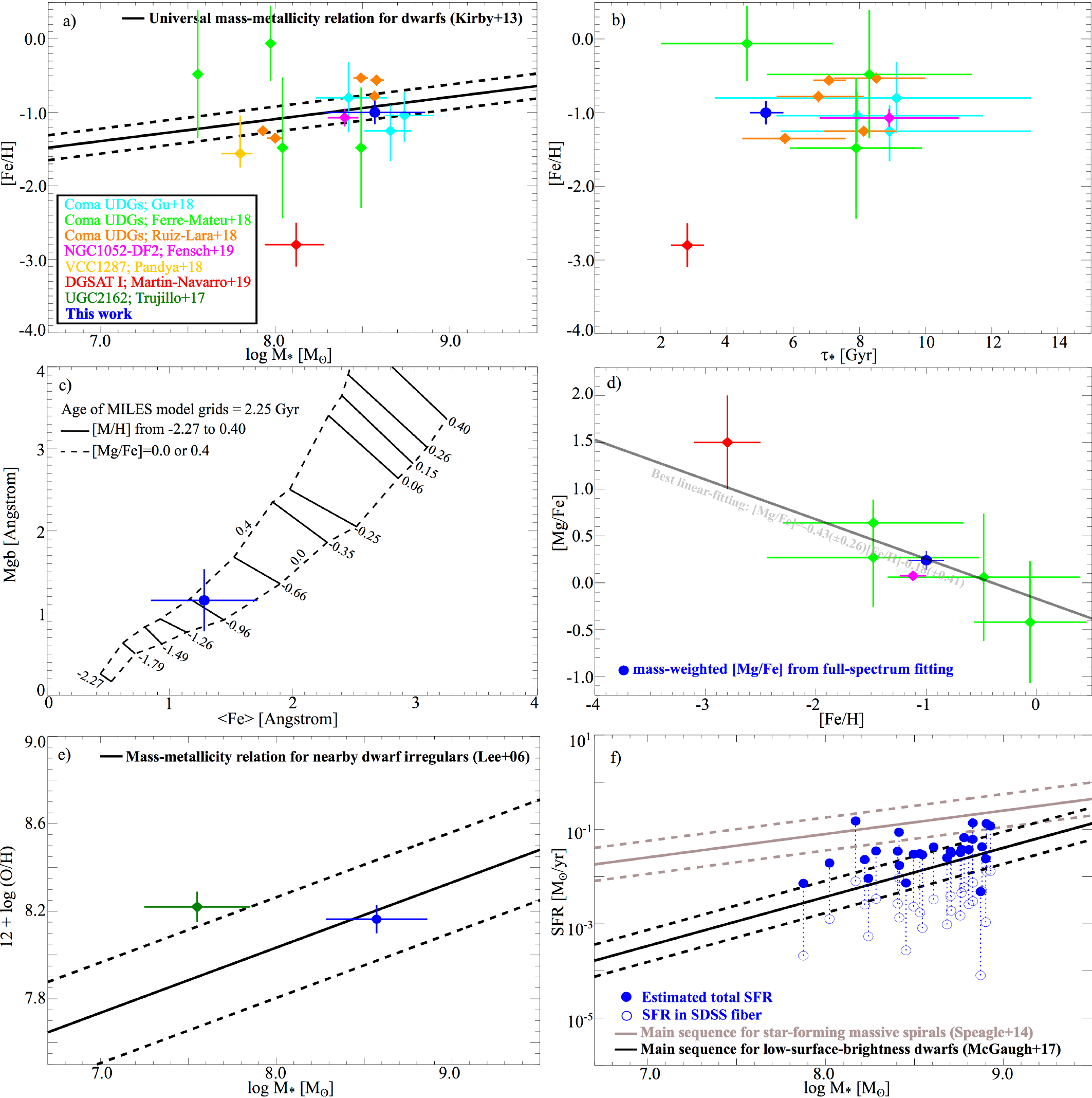}
\caption{Panel~a: stellar metallicities [Fe/H] versus stellar masses $M_*$ for UDGs, compared with the universal [Fe/H]-$M_*$ relation of nearby dwarf galaxies obtained from \cite{Kirby13}. Panel~b:~[Fe/H] versus stellar ages $\tau_*$ for UDGs. Panel~c:~Mgb vs. $\langle {\rm{Fe}} \rangle$ of our UDGs plotted onto the MILES model grids (the model age closest to the light-weighted $t_*\sim 2.2$~Gyr, i.e., 2.25~Gyr, is chosen). Panel~d:~[Mg/Fe] vs. [Fe/H] for UDGs (for our star-forming UDGs, we use the mass-weighted [Mg/Fe] value from the pPXF full-spectrum fitting); the best linear-fitting result for the [Mg/Fe]\--[Fe/H] relation of these UDGs is also shown. Panel~e:~oxygen abundance \Oabund\ as a function of $M_*$ for UDGs, compared with the mass-metallicity relation for nearby star-forming dwarfs obtained by \cite{Lee06}. Panel~f:~SFRs vs. $M_*$ for UDGs, compared with the main sequences of star-forming massive spirals (brown) and low-surface-brightness dwarfs (black) obtained by \cite{Speagle14} and \cite{McGaugh17}, respectively; the blue closed circles and open circles show the estimated total SFRs (i.e., upper-limits) and SFRs covered by the SDSS 3~arcsec fiber (the two SFRs of each UDG is linked by a dotted line), respectively. In the six panels, the blue color always denotes our UDGs in this work, while the cyan, light-green, orange, magenta, yellow, red, and dark-green diamonds denote the UDGs in previous literature of \cite{Gu18}, \cite{Ferre-Mateu18}, \cite{Ruiz-Lara18}, \cite{Fensch19}, \cite{Pandya18}, \cite{Martin-Navarro19}, and \cite{Trujillo17}, respectively (indicated in the inset of panel~a).}
\label{UDG_property}
\end{figure*}

{\textit{Stellar properties:}} We obtain the on-average mass-weighted total-metallicity [M/H]$=-0.82\pm 0.14$ and stellar age $t_*=5.2\pm0.5$~Gyr, as well as [$\alpha$/Fe]$\simeq 0.24\pm0.10$ (the \textsc{pPXF} fitting also gives the light-weighted [M/H]$\sim-0.93\pm0.17$, $t_*\sim2.2\pm0.7$~Gyr, and [$\alpha$/Fe]$\sim 0.27\pm0.11$). The mass-weighted iron-metallicity [Fe/H]$\simeq -1.00\pm0.16$ is estimated from [Fe/H]$\simeq$[M/H]-0.75[$\alpha$/Fe] \citep{Vazdekis15}.
%and light-weighted [Fe/H]$=-1.38\pm 0.17$ and $t_*=2.7\pm0.7$~Gyr. 
As explored in panel~a of Fig.~\ref{UDG_property}, similar to the mass-metallicity relations of UDGs in galaxy clusters/groups \citep{Gu18,Ferre-Mateu18,Fensch19,Ruiz-Lara18,Pandya18}, our UDGs follow (or located slightly lower than) the universal [Fe/H]$\--M_*$ relation of the nearby typical dwarf galaxies \citep{Kirby13}; in this sense, our UDGs are not particularly metal-poor or metal-rich for their stellar masses. Yet our UDGs in the low-density environments are younger than most of the member UDGs in galaxy clusters/groups \citep{Gu18,Ferre-Mateu18,Fensch19,Ruiz-Lara18} but older than the isolated UDG DGSAT~I \citep{Martin-Navarro19}, as shown in panel~b of Fig.~\ref{UDG_property}.

Since the information contained in relevant Mg- and Fe-sensitive features might be diluted by full-spectrum fitting, we also utilize the line strengths of Mgb, Fe5270, Fe5335, measured with the Lick/IDS index definitions of \cite{Worthey94}, to directly estimate [Mg/Fe]. Analogous to the method of \cite{Martin-Navarro19}, we plot the Mgb versus $\langle$Fe$\rangle =$(Fe5270+Fe5335)/2 of our UDGs onto the SSP model grids of MILES (we choose to plot the models with $t_*\sim 2.25$~Gyr, closest to the light-weighted age from the full-spectrum fitting), which have been broadened to match the resolution of the stacked spectrum, i.e., $\sigma_{\rm{SSP}}\simeq \sqrt{\sigma_{\rm{SDSS}}^2+\sigma_{\rm{los}}^2}$ (where the SDSS resolution $\sigma_{\rm{SDSS}}\simeq 2.76/2.355$~\AA~corresponds to $\sim 67$~km/s in 5140\--5365~\AA~covering Mgb, Fe5270, \& Fe5335, and $\sigma_{\rm{los}}$ is the dispersion of our UDGs; see below), as shown in panel~c of Fig.~\ref{UDG_property}. We interpolate the model grids and find [Mg/Fe]$\simeq 0.29\pm0.27$, similar to the light-weighted [Mg/Fe] from the full-spectrum fitting{\footnote{Hence, the full-spectrum fitting results can reveal the [Mg/Fe] of UDGs; hereafter, we always use the mass-weighted [Mg/Fe] from the full-spectrum fitting because that the light-weighted properties may be dominated by the youngest stars in our star-forming UDGs.}}. As shown in panel~d of Fig.~\ref{UDG_property}, our relatively-isolated UDGs have a lower [Mg/Fe] compared with the extremely-high [Mg/Fe]$\sim 1.5\pm0.5$ of DGSAT~I \citep{Martin-Navarro19}; yet it is similar to the [Mg/Fe] of the member UDGs in clusters/groups \citep{Ferre-Mateu18,Fensch19}. With these [Fe/H] vs. [Mg/Fe] data of UDGs in fields and clusters, we use a linear fitting to estimate the [Mg/Fe]\--[Fe/H] relation of UDGs, and derive [Mg/Fe]$\simeq-0.43(\pm0.26)$[Fe/H]$-0.18(\pm0.41)$.

In panel~d of Fig.~\ref{spectra}, we also show the on-average cumulative SFH of our UDGs using the regularized \textsc{pPXF} solution (black solid; using the 2nd-order regularization matrix $\bm B$); for comparison, we also show the SFH without regularization (dotted) and SFHs of applying the first- (dashed) and third-order (dot-dashed) $\bm B$ \citep[cf.][]{Boecker20}. The different regularization methods uniformly recover an extended SFH{\footnote{The different regularization methods also give similar [M/H], $t_*$, and [$\alpha$/Fe], considering their uncertainties.}}, lasting for more than 10~Gyr, similar to the extended SFHs of other UDGs in clusters and fields \citep{Ferre-Mateu18,Martin-Navarro19}. Following the regularized solution of applying the 2nd-order $\bm B$, we find that at the redshift $z\sim1.2$ and $\sim 0.2$ (corresponding to the lookback time $t\sim 8.6$~Gyr and $t\sim 2.1$~Gyr, respectively), our UDGs assembled their 50\% and 90\% stellar masses, respectively. 
%It should be noted that the regularized \textsc{pPXF} solution may not completely reflect the true SFH of the galaxies, which is likely to be stochastic and variable over short timescales \citep{Cappellari17}.

To estimate the on-average stellar velocity dispersion $\sigma_{\rm{los}}$ of our UDGs, we set the additive polynomials of the 12th order and multiplicative polynomials of the 14th order \citep{Fensch19}, and fit the stacked spectrum again. We find $\sigma_{\rm{los}}\simeq 54\pm12$~km/s for our UDGs{\footnote{We have used the mock spectra with the different input stellar dispersions but same S/N ($\sim 30$) of our stacked spectrum, and found that the pPXF fitting can well recover a dispersion $>35$~km/s; for $\sigma_{\rm{los}}<35$~km/s, the fitting slightly under-estimates (but still in $1\sigma$ uncertainty range) the input dispersions \citep[cf. also][]{Guerou17,Cappellari17}.}}, comparable to the high dispersion of DGSAT~I \citep{Martin-Navarro19}. However, note that $\sigma_{\rm{los}}$ only suggests the central stellar dispersion of our UDGs, since the SDSS fiber primarily targets at the central regions of our UDGs.

{\textit{Gas-phase properties:}} After subtracting the best-fit stellar models from the stacked spectrum, we then use a Gaussian profile to fit each emission line carefully and estimate the on-average gas-phase metallicity. Since the [O\,{\sc ii}] lines are not covered by the wavelength range of the stacked spectrum, we use two additional powerful diagnostics, i.e., N2S2H$\alpha$ defined by \cite{Dopita16} and O3N2 described in \cite{Pettini04}, to estimate the oxygen abundance, respectively. The former diagnostic makes use of the flux ratios of \Nii$\lambda6584$/H$\alpha$ and \Nii$\lambda6584$/\Sii$\lambda\lambda$6717,31, while the latter one applies \Oiii$\lambda5007$/H$\beta$ and \Nii$\lambda6584$/H$\alpha$, to determine the O/H ratio. By the uses of emission lines located close together in wavelength, the two diagnostics are actually independent of the internal extinction. We obtain \Oabund$\simeq 7.94\pm0.10$ (N2S2H$\alpha$) and $\simeq 8.38\pm0.07$ (O3N2) for our UDGs, and treat the mean \Oabund\ value from the two diagnostics as the final on-average metallicity. As shown in panel~e of Fig.~\ref{UDG_property}, we find that, different from the relatively-high oxygen abundance of the star-forming UDG UGC~2162 \citep[$g-r\simeq 0.45$, \Oabund=$8.22\pm 0.07$;][]{Trujillo17}, the on-average oxygen abundance of our UDGs follow (within $1\sigma$ uncertainty) the  \Oabund\--$M_*$ relation of the nearby star-forming dwarf galaxies \citep{Lee06}, confirming again that our UDGs in the low-density environments are not particularly metal-poor or metal-rich.

In order to assess the star-formation rate (SFR) of each UDG, we impose to fit the spectrum of each UDG with the optimal stellar template, and derive the H$\alpha$ emission line flux covered by the SDSS $3''$ fiber from the residual spectrum. The SFR in fiber aperture, SFR$_{\rm{fiber}}$, is obtained by adopting the H$\alpha$ luminosity-SFR relation of \cite{Kennicutt94}. We also estimate the total SFR of each UDG by using the ratio of luminosities of the region covered by the fiber and entire galaxy; note that the total SFR is actually the upper-limit, since the SDSS fiber primarily targets at the central regions of our UDGs, while the star formation in a dwarf galaxy is usually concentrated at the central region. As shown in panel~f of Fig.~\ref{UDG_property}, on the SFR versus $M_*$ diagram, our UDGs are distributed lower than the extrapolated star-forming main sequence from the massive spirals \citep[brown;][]{Speagle14}, but plausibly follow (or be slightly lower than) the main sequence of the low-surface-brightness dwarf galaxies \citep[black;][]{McGaugh17}, suggesting a possible lower star-formation efficiency (i.e., low SFR/H$_2$) or HI-to-H$_2$ ratio in these UDGs.

%%%%%%%%%%%%%%%%%%%%%%%%%%%%%%%%%%%%%%%%%%%%%
\section{Discussion}\label{sec:4}

In this work, for our small UDG sample including 28 members, we used the bootstrap methodology to estimate the uncertainty of each on-average property. We randomly sampled the spectra of the 28 UDGs with replacement for 1,000 times; in each sampling, we stack the 28 sampled spectra and fit the stacked spectrum following the steps described in section~\ref{sec:3}, and thus obtain 1,000 numbers of $t_*$, [M/H], [$\alpha$/Fe], line indices, emission-line fluxes, and dispersions, etc. For $t_*$, [M/H], line indices, and emission-line fluxes, their standard deviations $\sigma_{\rm{std}}$ are treated as the uncertainties. For [$\alpha$/Fe], since we linearly interpolated the SSP models between [$\alpha$/Fe]=0 and 0.4~dex with a spacing of 0.1~dex, we included an additional error of $\sim0.1$~dex which is the maximum [$\alpha$/Fe] uncertainty possibly introduced by interpolation, i.e., the [$\alpha$/Fe] uncertainty~$\simeq \sqrt{\sigma_{\rm{std}}^2+0.1^2}$. For $\sigma_{\rm{los}}$, we included the average redshift uncertainty of the 28 UDGs, i.e., $\sigma_z\sim 2.4\times10^{-5}$ corresponding to a dispersion error of $\sim 7$~km/s, therefore, the stellar dispersion uncertainty~$\simeq \sqrt{\sigma_{\rm{std}}^2+7^2}$.

Since our UDGs have very-extended SFH and assembled their 50\% and 90\% stellar masses at $z\sim 1.2$ and $\sim0.2$ respectively, it may reject the current failed $L^*$ UDG formation model \citep{Yozin15}, where UDGs should be quenched at $z\gtrsim 2$. Besides, the tidal interaction with massive galaxies is also very unlikely to be the formation mechanism for our relatively-isolated UDGs.

The light-weighted stellar age ($t_*\sim2.2\pm0.7$~Gyr) from the \textsc{pPXF} full-spectrum fitting is smaller than the mass-weighted age ($t_*\sim5.2\pm0.5$~Gyr); as shown in panel~d of Fig.~\ref{spectra}, 30\% light is contributed by the recently-formed stars with $t_*<1$~Gyr. These suggest that the light-weighted metallicity values should be significantly affected by the youngest stars. However, the light-weighted metallicities (including [M/H] and [$\alpha$/Fe]) are comparable to the mass-weighted metallicities; it probably indicates that the metal-rich outflows or metal-poor inflows reduced the metallicities produced by the previous generations of stellar populations, since the recently-formed stars in our UDGs do not show significantly-higher metallicities than the underlying old stellar populations. The results may be compatible with the current stellar-feedback model of \cite{Chan18} or high-spin model of \cite{Rong17a}, which can predict the outflows or inflows during the formation of isolated UDGs as well as present-day star-forming UDGs with the low specific SFRs and stellar ages/metallicities similar to our results. 

However, it is also worth to note that, our isolated UDGs are not particularly metal-poor/rich for their stellar masses, as their [Fe/H]-$M_*$ and \Oabund-$M_*$ relations follow the mass-metallicity relations of typical dwarfs; it suggests that the feedback-driven outflows in UDGs were not particularly stronger than those in the typical dwarf counterparts \citep[e.g.,][]{Spitoni10}.

Yet, note also that our UDGs are relatively-isolated, star-forming, and thus represent the relatively-bright side of UDG populations as shown in Fig.~\ref{scalerelation}; therefore, there may be a systematic property bias of our star-forming UDGs from that of the entire UDG population.

Finally, for panel~d of Fig.~\ref{UDG_property}, we indicate that the [Mg/Fe] based on the different SSP models may be different, particularly for the low-metallicity cases; therefore, in order to obtain a more accurate [Mg/Fe] vs. [Fe/H] relation, the [Mg/Fe] of NGC~1052-DF2 obtained from the SSP models of \cite{Thomas11} (TMJ11) should be estimated again with the MILES SSP models \citep{Fensch19}. Using the line indices values of NGC~1052-DF2 given by \cite{Fensch19}, we find that TMJ11 gives a lower [Mg/Fe], compared with the MILES models (with a difference of $\Delta$[Mg/Fe]$\sim 0.3$~dex). After the revision, we obtain a corrected relation of [Mg/Fe]$\simeq-0.43(\pm0.26)$[Fe/H]$-0.14(\pm0.40)$ for UDGs.

%%%%%%%%%%%%%%%%%%%%%%%%%%%%%%%%%%%%%%%%%%%%%%%%%%%%%%%%%%%%%%%%%%%%%%%%%%%%%%%%
\acknowledgments

We thank the referee for their comments and suggestions, thank I. Martin-Navarro and K. Fahrion for their helpful discussions, and thank Qi Guo, Zheng Zheng, Hui-Jie Hu, and Xiaoyu Dong for their helps. Y.R. acknowledges funding supports from FONDECYT Postdoctoral Fellowship Project No.~3190354 and NSFC grant No.\,11703037. T.H.P. acknowledges support through FONDECYT Regular project 1161817 and CONICYT project Basal AFB-170002. H.X.Z. acknowledges support from the CAS Pioneer Hundred Talents Program and the NSFC grant 11421303. This research is/was (partially) based on data from the MILES project.

%%%%%%%%%%%%%%%%%%%%%%%%%%%%%%%%%%%%%
\vspace{5mm}

%%%%%%%%%%%%%%%%%%%%%%%%%%%%%%%%%%%%%

%%%%%%%%%%%%%%%%%%%%%%%%%%%%%%%%%%%%%%%%%%%%%%%%%%


\begin{thebibliography}{}

%\bibitem[\protect\citeauthoryear{Andrews \& Martini}{2013}]{Andrews13} Andrews, B. H., \& Martini, P. 2013, ApJ, 765, 140
%\bibitem[\protect\citeauthoryear{Asplund et al.}{2009}]{Asplund09} Asplund, M., Grevesse, N., Sauval, A. J., \& Scott, P. 2009, ARA\&A, 47, 481
%\bibitem[\protect\citeauthoryear{Barkana \& Loeb}{2001}]{Barkana01} Barkana, R., Loeb, A. 2001, Physics Reports, 349, 125
%\bibitem[\protect\citeauthoryear{Bell et al.}{2003}]{Bell03} Bell E. F., McIntosh D. H., Katz N. \& Weinberg M. D. 2003, ApJS, 149, 289
%\bibitem[\protect\citeauthoryear{Beasley et al.}{2016}]{Beasley16} Beasley, M. A., Romanowsky, A. J., Pota, V., Navarro, I. M., Martinez Delgado, D., Neyer, F., Deich, A. L. 2016, ApJ, 819L, 20
%\bibitem[\protect\citeauthoryear{Beckmann et al.}{2017}]{Beckmann17} Beckmann, R. S., et al. 2017, MNRAS, 472, 949
\bibitem[\protect\citeauthoryear{Bennett et al.}{2014}]{Bennett14} Bennett, C. L., Larson, D., Weiland, J. L., Hinshaw, G. 2014, ApJ, 794, 135
%\bibitem[\protect\citeauthoryear{Bian et al.}{2018}]{Bian18} Bian, F., Kewley, L. J., Dopita, M. A. 2018, ApJ, 859, 175
\bibitem[\protect\citeauthoryear{Boecker et al.}{2020}]{Boecker20} Boecker, A., Leaman, R., van de Ven, G., Norris, M. A., Mackereth, T., Crain, R. A. 2020, MNRAS, 491, 823
%\bibitem[\protect\citeauthoryear{Calzetti et al.}{2000}]{Calzetti00} Calzetti, D., Armus, L., Bohlin, R. C., Kinney, A. L., Koornneef, J. 2000, ApJ, 533, 682
\bibitem[\protect\citeauthoryear{Cappellari}{2017}]{Cappellari17} Cappellari, M. 2017, MNRAS, 466, 798
%\bibitem[\protect\citeauthoryear{Cappellari \& Cappellari}{2004}]{Cappellari04} Cappellari, M., \& Emsellem, E. 2004, PASP, 116, 138
%\bibitem[\protect\citeauthoryear{Cardelli et al.}{1998}]{Cardelli98} Cardelli, J. A., Clayton, G. C., \& Mathis, J. S. 1989, ApJ, 345, 245
\bibitem[\protect\citeauthoryear{Chan et al.}{2018}]{Chan18} Chan, T. K., Keres, D., Wetzel, A., Hopkins, P. F., Faucher-Gigu\`ere, C. -A., El-Badry, K., Garrison-Kimmel, S., Boylan-Kolchin, M. MNRAS, 2018, 478, 906
%\bibitem[\protect\citeauthoryear{Chilingarian et al.}{2019}]{Chilingarian19} Chilingarian, I. V., Afanasiev, A. V., Grishin, K. A., Fabricant, D., Moran, S. 2019, ApJ, 884, 79
\bibitem[\protect\citeauthoryear{Christensen et al.}{2018}]{Christensen18} Christensen, C. R., Dav\'e, R., Brooks, A., Quinn, T., Shen, S. 2018, ApJ, 867, 142
%\bibitem[\protect\citeauthoryear{Conselice et al.}{2018}]{Conselice18} Conselice, C. J. 2018, RNAAS, 2, 43
%\bibitem[\protect\citeauthoryear{Curti et al.}{2017}]{Curti17} Curti M., Cresci G., Mannucci F., Marconi A., Maiolino R., Esposito S., 2017, MNRAS, 465, 1384
%\bibitem[\protect\citeauthoryear{de Naray et al.}{2004}]{Naray04} de Naray R., McGaugh S., de Blok W. 2004, MNRAS, 355, 887
%\bibitem[\protect\citeauthoryear{Dekel \& Woo}{2003}]{Dekel03} Dekel, A., Woo., J. 2003, MNRAS, 344, 1131
%\bibitem[\protect\citeauthoryear{Di Cintio et al.}{2017}]{DiCintio17} Di Cintio, A., Brook, C. B., Dutton, A. A., Maccio`, A. V., Obreja, A., Dekel, A. 2017, MNRAS, 466L, 1
\bibitem[\protect\citeauthoryear{Dopita et al.}{2016}]{Dopita16} Dopita, M. A., Kewley, L. J., Sutherland, R. S., Nicholls, D. C. 2016, Astrophysics and Space Science, 361, 61
%\bibitem[\protect\citeauthoryear{Du et al.}{2017}]{Du17} Du, W., Wu, H., Zhu, Y., Zheng, W., Filippenko, A. V. 2017, ApJ, 837, 152
\bibitem[\protect\citeauthoryear{Eigenthaler et al.}{2018}]{Eigenthaler18} Eigenthaler, P., et al. 2018, ApJ, 855, 142
%\bibitem[\protect\citeauthoryear{Emsellem et al.}{2019}]{Emsellem19} Emsellem, E., et al., 2019, A\&A, 625, 76
\bibitem[\protect\citeauthoryear{Fahrion et al.}{2019}]{Fahrion19} Fahrion, K., et al. 2019, A\&A, 628, 92
%\bibitem[\protect\citeauthoryear{Feldmann et al.}{2012}]{Feldmann12} Feldmann, R., Gnedin, N. Y., Kravtsov, A. V. 2012, ApJ, 747, 124
\bibitem[\protect\citeauthoryear{Fensch et al.}{2019}]{Fensch19} Fensch, J., et al. 2019, A\&A, 625, 77
\bibitem[\protect\citeauthoryear{Ferrarese et al.}{2012}]{Ferrarese12} Ferrarese, L., et al. 2012, ApJS, 200, 4
\bibitem[\protect\citeauthoryear{Ferr\'e-Mateu et al.}{2018}]{Ferre-Mateu18} Ferr\'e-Mateu, A., et al. 2018, MNRAS, 479, 4891
%\bibitem[\protect\citeauthoryear{Ferr\'e-Mateu et al.}{2013}]{Ferre-Mateu13} Ferr\'e-Mateu, A., et al. 2013, MNRAS, 431, 440
\bibitem[\protect\citeauthoryear{Fitzpatrick}{1999}]{Fitzpatrick99} Fitzpatrick, E. L. 1999, PASP, 111, 63
%\bibitem[\protect\citeauthoryear{Glover \& Mac Low}{2011}]{Glover11} Glover, S. C. O., Mac Low, M.-M. 2011, MNRAS, 412, 337
%\bibitem[\protect\citeauthoryear{Gu\'erou et al.}{2017}]{Guerou17} Gu\'erou, A., et al. 2017, A\&A, 608, 5
\bibitem[\protect\citeauthoryear{Guo et al.}{2020}]{Guo20} Guo, Q., et al., 2020, Nature~Astronomy, 4, 246
%\bibitem[\protect\citeauthoryear{Ginolfi et al.}{2019}]{Ginolfi19} Ginolfi, M., Hunt, L. K., Tortora, C., Schneider, R., Cresci, G. 2019, eprint arXiv:~190706654
%\bibitem[\protect\citeauthoryear{Giovanelli}{2007}]{Giovanelli07} Giovanelli R. 2007, Nuovo Cimento B, 122, 1097
%\bibitem[\protect\citeauthoryear{Girardi et al.}{2000}]{Girardi00} Girardi, L. Bressan, A., Bertelli, G., Chiosi, C. 2000, A\&AS, 141, 371
%\bibitem[\protect\citeauthoryear{Glover \& Mac Low}{2011}]{Glover11} Glover, S. C. O., Mac Low, M. -M. 2011, MNRAS, 412, 337
\bibitem[\protect\citeauthoryear{Gu et al.}{2018}]{Gu18} Gu, M., et al. 2018, ApJ, 859, 37
\bibitem[\protect\citeauthoryear{Gu\'erou et al.}{2017}]{Guerou17} Gu\'erou, A., et al. 2017, A\&A, 608, 5
%\bibitem[\protect\citeauthoryear{He et al.}{2019}]{He19} He, M., Wu, H., Du, W., Wicker, J., Zhao, P., Lei, F., Liu, J. 2019, ApJ, 880, 30
\bibitem[\protect\citeauthoryear{Ho et al.}{2011}]{Ho11} Ho, L. C., Li, Z.-Y., Barth, A. J., Seigar, M. S., Peng, C. Y. 2011, ApJS, 197, 21
%\bibitem[\protect\citeauthoryear{Janssens et al.}{2017}]{Janssens17} Janssens, S., Abraham, R., Brodie, J., Forbes, D., Romanowsky, A. J., van Dokkum, P. 2017, ApJ, 839L, 17
%\bibitem[\protect\citeauthoryear{Johnston et al.}{2020}]{Johnston20} Johnston, E., et al. 2020, eprint arXiv:~2005.01532
%\bibitem[\protect\citeauthoryear{Kadowaki et al.}{2017}]{Kadowaki17} Kadowaki, J., Zaritsky, D., Donnerstein, R. L. 2017, ApJ, 838L, 21
%\bibitem[\protect\citeauthoryear{Kennicutt \& Evans}{2012}]{Kennicutt12} Kennicutt, R. C., Evans, N. J. 2012, ARA\&A, 50 531
\bibitem[\protect\citeauthoryear{Kennicutt}{1994}]{Kennicutt94} Kennicutt, R. C., Tamblyn, P., Congdon, C.W. 1994, ApJ, 435, 22
%\bibitem[\protect\citeauthoryear{Kewley et al.}{2006}]{Kewley06} Kewley, L. J.,  Geller, M. J.,  Barton, E. J. AJ, 2006, 131, 2004
%\bibitem[\protect\citeauthoryear{Kewley et al.}{2010}]{Kewley10} Kewley, L. J., Rupke, D., Zahid, H. J., Geller, M. J., Barton, E. J. 2010, ApJ, 721L, 48
\bibitem[\protect\citeauthoryear{Kirby et al.}{2013}]{Kirby13} Kirby, E. N., Cohen, J. G., Guhathakurta, P., Cheng, L., Bullock, J. S., Gallazzi, A. 2013, ApJ, 779, 102
%\bibitem[\protect\citeauthoryear{Lee et al.}{2003}]{Lee03} Lee H., McCall M. L.,Kingsburgh R. L., Ross R., Stevenson C. C. 2003, AJ, 125, 146
\bibitem[\protect\citeauthoryear{Lee et al.}{2006}]{Lee06} Lee, H., Skillman, E. D., Cannon, J. M., et al. 2006, ApJ, 647, 970
\bibitem[\protect\citeauthoryear{Leisman et al.}{2017}]{Leisman17} Leisman, L., et al. 2017, ApJ, 842, 133
%\bibitem[\protect\citeauthoryear{Li et al.}{2017}]{Li17} Li, H., et al. 2017, ApJ, 838, 77
%\bibitem[\protect\citeauthoryear{Liu et al.}{2016}]{Liu16} Liu, Y., et al. 2016, ApJ, 818, 179
\bibitem[\protect\citeauthoryear{Mart\'in-Navarro et al.}{2019}]{Martin-Navarro19} Mart\'in-Navarro, I., et al. 2019, MNRAS, 484, 3425
\bibitem[\protect\citeauthoryear{McDermid et al.}{2015}]{McDermid15} McDermid, R. M., et al. 2015, MNRAS, 448, 3484
%\bibitem[\protect\citeauthoryear{McConnachie}{2012}]{McConnachie12} McConnachie, A. W. 2012, AJ, 144, 4
\bibitem[\protect\citeauthoryear{McGaugh et al.}{2017}]{McGaugh17} McGaugh, S. S., Schombert, J. M., Lelli, F. 2017, ApJ, 851, 22
\bibitem[\protect\citeauthoryear{Mihos et al.}{2015}]{Mihos15} Mihos, J. C., et al. 2015, ApJ, 809L, 21
%\bibitem[\protect\citeauthoryear{Montuori et al.}{2010}]{Montuori10} Montuori, M., Di Matteo, P., Lehnert, M. D., Combes, F., Semelin, B. 2010, A\&A, 518, 56
\bibitem[\protect\citeauthoryear{Pandya et al.}{2018}]{Pandya18} Pandya, V., et al. 2018, ApJ, 858, 29
%\bibitem[\protect\citeauthoryear{Peng et al.}{2002}]{Peng02} Peng, C. Y., Ho, L. C., Impey, C. D., Rix, H.-W. 2002, AJ, 124, 266 
\bibitem[\protect\citeauthoryear{Peng et al.}{2010}]{Peng10} Peng, C. Y., Ho, L. C., Impey, C. D., Rix, H.-W. 2010, AJ, 139, 2097
%\bibitem[\protect\citeauthoryear{Peng et al.}{2012}]{Peng12} Peng, Y.-J., Lilly, S. J., Renzini, A., Carollo, M. 2012, ApJ, 757, 4
\bibitem[\protect\citeauthoryear{Pettini \& Pagel}{2004}]{Pettini04} Pettini, M. \& Pagel, B. E. J. 2004, MNRAS, 348, L59
\bibitem[\protect\citeauthoryear{Pietrinferni et al.}{2006}]{Pietrinferni06} Pietrinferni, A., Cassisi, S., Salaris, M., Castelli, F. 2006, ApJ, 642, 797
%\bibitem[\protect\citeauthoryear{Pipino \& Matteucci}{2004}]{Pipino04} Pipino, A., Matteucci, F. 2004, MNRAS, 347, 968
%\bibitem[\protect\citeauthoryear{Press et al.}{1992}]{Press92} Press W. H., Teukolsky S. A., Vetterling W. T., Flannery B. P. 1992, Numerical recipes in FORTRAN. The art of scientific computing. Cambridge: University Press, |c1992, 2nd ed.
%\bibitem[\protect\citeauthoryear{Rom\'an \& Trujillo}{2017}]{Roman17} Rom\'an, J., Trujillo, I. 2017, MNRAS, 468, 4039
%\bibitem[\protect\citeauthoryear{Rom\'an \& Trujillo}{2017b}]{Roman17b} Rom\'an, J., Trujillo, I. 2017b, MNRAS, 468, 703
%\bibitem[\protect\citeauthoryear{Rong et al.}{2015a}]{Rong15a} Rong, Y., Yi, S.-X., Zhang, S.-N., Tu, H. 2015a, MNRAS, 451, 2536
%\bibitem[\protect\citeauthoryear{Rong et al.}{2015b}]{Rong15b} Rong, Y., Zhang, S.-N., Liao, J.-Y. 2015b, MNRAS, 453, 1577
%\bibitem[\protect\citeauthoryear{Rong et al.}{2016}]{Rong16} Rong, Y., Liu, Y., Zhang, S.-N. 2016, MNRAS, 455, 2267
\bibitem[\protect\citeauthoryear{Rong et al.}{2017a}]{Rong17a} Rong, Y., Guo, Q., Gao, L., Liao, S., Xie, L., Puzia, T. H., Sun, S., Pan, J. 2017a, MNRAS, 470, 4231
\bibitem[\protect\citeauthoryear{Rong et al.}{2018}]{Rong18a} Rong, Y., Li, H., Wang, J., et al. 2018, MNRAS, 477, 230
%\bibitem[\protect\citeauthoryear{Rong et al.}{2018b}]{Rong18b} Rong, Y., et al. 2018b, eprint arXiv: 1806.10149
\bibitem[\protect\citeauthoryear{Rong et al.}{2017b}]{Rong17b} Rong, Y., Jing, Y., Gao, L., Guo, Q., Wang, J., Sun, S., Wang, L., Pan, J. 2017b, MNRAS, 471L, 36
\bibitem[\protect\citeauthoryear{Rong et al.}{2019a}]{Rong19a} Rong, Y., et al., 2019a, ApJ, 883, 56
\bibitem[\protect\citeauthoryear{Rong et al.}{2019b}]{Rong19b} Rong, Y., et al., 2019b, eprint arXiv: 1907.10079
\bibitem[\protect\citeauthoryear{Rong et al.}{2020}]{Rong20} Rong, Y., Mancera Pi\~na, P. E., Tempel, E., Puzia, T. H., De Rijcke, S. 2020, eprint arXiv:~2007.06593
\bibitem[\protect\citeauthoryear{Rupke et al.}{2010}]{Rupke10} Rupke, D. S. N., Kewley, L. J., Barnes, J. E. 2010, ApJ, 710L, 156
%\bibitem[\protect\citeauthoryear{Rupke et al.}{2008}]{Rupke08} Rupke, D. S. N., Veilleux, S., Baker, A. J. 2008, ApJ, 674, 172
\bibitem[\protect\citeauthoryear{Ruiz-Lara et al.}{2018}]{Ruiz-Lara18} Ruiz-Lara, T., et al. 2018, MNRAS, 478, 2034
%\bibitem[\protect\citeauthoryear{Salpeter}{1955}]{Salpeter55} Salpeter, E. E. 1955, ApJ, 121, 161
%\bibitem[\protect\citeauthoryear{S\'anchez Almeida et al.}{2018}]{Sanchez-Almeida18} S\'anchez Almeida, J., Olmo-Garc\'ia, A., Elmegreen, B. G., Elmegreen, D. M., Filho, M.; Mu\~noz-Tu\~n\'on, C., P\'erez-Montero, E., Rom\'an, J. 2018, ApJ, 869, 40
%\bibitem[\protect\citeauthoryear{S\'anchez-Bl\'azquez et al.}{2006}]{Sanchez06} S\'anchez-Bl\'azquez, P., et al. 2006, MNRAS, 371, 703
%\bibitem[\protect\citeauthoryear{Salaris et al.}{1993}]{Salaris93} Salaris, M., Chieffi, A., Straniero, O. 1993, ApJ, 414, 580
\bibitem[\protect\citeauthoryear{Saulder et al.}{2016}]{Saulder16} Saulder, C., van Kampen, E., Chilingarian, I. V., Mieske, S., Zeilinger, W. W. 2016, A\&A, 596, 14
%\bibitem[\protect\citeauthoryear{Schlegel et al.}{1998}]{Schlegel98} Schlegel D. J., Finkbeiner D. P., Davis M., 1998, ApJ, 500, 525
%\bibitem[\protect\citeauthoryear{Shi et al.}{2014}]{Shi14} Shi, Y., et al., 2014, Nature, 514, 335
%\bibitem[\protect\citeauthoryear{Shi et al.}{2016}]{Shi16} Shi, Y., Wang, J., Zhang, Z.-Y., Gao, Y., Hao, C.-N. Xia, X.-Y., Gu, Q., 2016, Nature Communications, 7, 13789
\bibitem[\protect\citeauthoryear{Simard et al.}{2011}]{Simard11} Simard, L., Mendel, J. T., Patton, D. R., Ellison, S. L., McConnachie, A. W. 2011, ApJS, 196, 11
%\bibitem[\protect\citeauthoryear{Singh et al.}{2019}]{Singh19} Singh, P. R., Zaritsky, D., Donnerstein, R., Spekkens, K. 2019, AJ, 157, 212
\bibitem[\protect\citeauthoryear{Speagle et al.}{2014}]{Speagle14} Speagle, J. S., Steinhardt, C. L., Capak, P. L., Silverman, J. D. 2014, ApJS, 214, 15
%\bibitem[\protect\citeauthoryear{Spekkens \& Karunakaran}{2018}]{Spekkens18} Spekkens, K., Karunakaran, A. 2018, ApJ, 855, 28
\bibitem[\protect\citeauthoryear{Spitoni et al.}{2010}]{Spitoni10} Spitoni,E., Calura, F., Matteucci, F., Recchi, S. 2010, A\&A, 514, 73
%\bibitem[\protect\citeauthoryear{Thomas et al.}{1999}]{Thomas99} Thomas, D., Greggio, L., Bender, R. 1999, MNRAS, 302, 537
%\bibitem[\protect\citeauthoryear{Thomas et al.}{2003}]{Thomas03} Thomas, D., Maraston, C., Bender, R. 2003, MNRAS, 339, 89
\bibitem[\protect\citeauthoryear{Thomas et al.}{2005}]{Thomas05} Thomas, D., Maraston, C., Bender, R., Mendes de Oliveira, C. 2005, ApJ, 621, 673
\bibitem[\protect\citeauthoryear{Thomas et al.}{2011}]{Thomas11} Thomas, D., Maraston, C., Johansson, J. 2011, MNRAS, 412, 2183
%\bibitem[\protect\citeauthoryear{Tremonti et al.}{2004}]{Tremonti04} Tremonti, C. A., et al. 2004, ApJ, 613, 898
\bibitem[\protect\citeauthoryear{Trujillo et al.}{2017}]{Trujillo17} Trujillo, I., Roman, J., Filho, M., S\'anchez Almeida, J. 2017, ApJ, 836, 191
%\bibitem[\protect\citeauthoryear{Valdes et al.}{2004}]{Valdes04} Valdes, F., Gupta, R., Rose, J. A., Singh, H. P., Bell, D. J. 2004, ApJS, 152, 251
\bibitem[\protect\citeauthoryear{van Dokkum et al.}{2015}]{vanDokkum15} van Dokkum, P. G., Abraham, R., Merritt, A., Zhang, J., Geha, M., Conroy, C. 2015, ApJ, 798L, 45
\bibitem[\protect\citeauthoryear{van Dokkum et al.}{2016}]{vanDokkum16} van Dokkum, P. G., et al. 2016, ApJL, 828, 6
\bibitem[\protect\citeauthoryear{van Dokkum et al.}{2018}]{vanDokkum18} van Dokkum, P. G., et al. 2018, Nature, 555, 629
%\bibitem[\protect\citeauthoryear{van Dokkum et al.}{2019}]{vanDokkum19} van Dokkum, P. G., Danieli, S., Abraham, R., Conroy, C., Romanowsky, A. J. 2019, ApJ, 874L, 5
%\bibitem[\protect\citeauthoryear{van der Burg et al.}{2017}]{vanderBurg17} van der Burg, R. F. J., et al. 2017, A\&A, 607, 79
%\bibitem[\protect\citeauthoryear{Vazdekis et al.}{2010}]{Vazdekis10} Vazdekis, A., S\'anchez-Bl\'azquez, P., Falc\'on-Barroso, J., Cenarro, A. J., Beasley, M. A., Cardiel, N., Gorgas, J., Peletier, R. F. 2010, MNRAS, 404, 1639
\bibitem[\protect\citeauthoryear{Vazdekis et al.}{2015}]{Vazdekis15} Vazdekis, A., et al. 2015, MNRAS, 449, 1177
%\bibitem[\protect\citeauthoryear{Wolf et al.}{2010}]{Wolf10} Wolf, J., Martinez G. D., Bullock J. S., Kaplinghat M., Geha M.,
%Mu\~noz R. R., Simon J. D., Avedo F. F., 2010, MNRAS, 406, 1220
%\bibitem[\protect\citeauthoryear{Wolfire et al.}{2008}]{Wolfire08} Wolfire, M. G., Tielens, A. G. G. M., Hollenbach, D., Kaufman, M. J. 2008, ApJ, 680, 384
%\bibitem[\protect\citeauthoryear{Worthey et al.}{1992}]{Worthey92} Worthey, G., Faber, S. M. Gonzalez, J. J. 1992, ApJ, 398, 69
\bibitem[\protect\citeauthoryear{Worthey et al.}{1994}]{Worthey94} Worthey, G., Faber, S. M. Gonzalez, J. J., Burstein, D. 1994, ApJS, 94, 687
\bibitem[\protect\citeauthoryear{Yagi et al.}{2016}]{Yagi16} Yagi, M., Koda, J., Komiyama, Y., Yamanoi, H. 2016, ApJS, 225, 11
\bibitem[\protect\citeauthoryear{Yozin \& Bekki}{2015}]{Yozin15} Yozin, C., Bekki, K. 2015, MNRAS, 452, 937

\end{thebibliography}
\end{document}